\documentclass{svproc}

\usepackage{url}

\usepackage{graphicx}
\usepackage{amsmath}
\usepackage{amssymb}
\usepackage{booktabs}

\begin{document}
\mainmatter

\title{Agentic AI for Clinical Urgency Mapping and Queue Optimization in High-Volume Outpatient Departments: A Simulation-Based Evaluation}

\titlerunning{Agentic AI for OPD Urgency Mapping}

\author{Ravish Gupta\inst{1} \and
Saket Kumar\inst{2} \and
Maulik Dang\inst{3}}

\authorrunning{Gupta et al.}

\institute{
Lead Software Engineer, BigCommerce; IEEE Senior Member\\
\email{ravishgupta@ieee.org}
\and
University at Buffalo, The State University of New York, Buffalo, NY, USA\\
\email{saketkmr.dev@gmail.com}
\and
Senior Software Engineer, Amazon\\
\email{dangmaulik@gmail.com}
}

\maketitle

\begin{abstract}
Outpatient departments (OPDs) in Indian public hospitals face severe overcrowding, with daily volumes reaching 200--8,000 patients~\cite{aiims2020annual}. The prevailing First-Come-First-Served (FCFS) token system treats all patients equally regardless of clinical urgency, leading to dangerous delays for critical cases. We present an agentic AI framework integrating six components: voice-based multilingual symptom capture (modeled), LLM-powered severity prediction, load-aware physician assignment, adaptive queue optimization with urgency drift detection, a multi-objective orchestrator, and a Patient Memory System for longitudinal context-aware triage. Evaluated through discrete-event simulation of a District Hospital in Jabalpur (Madhya Pradesh) with 368 synthetic patients over 30 runs, the framework achieves 94.2\% critical patients seen within 10 minutes (vs.~30.8\% under FCFS), detects $\sim$236 simulated urgency drift events per session (modeled via stochastic deterioration probabilities), identifies $\sim$11.9 additional hidden-critical cases via patient memory, and recomposes queue urgency distribution from 13/36/158/161 (Critical/High/Medium/Low) to $\sim$25/178/115/50 through continuous reassessment, while maintaining comparable throughput ($\sim$40.4 patients/hour).
\end{abstract}

\section{Introduction}

India's public healthcare system serves over 1.4 billion people, with outpatient departments (OPDs) functioning as the primary point of contact for the vast majority seeking medical care. Government hospitals in tier-2 cities like Jabalpur, Madhya Pradesh, routinely handle 200--400 OPD patients per session with limited specialist physicians, creating chronic overcrowding and extended wait times that average 37--75 minutes nationally~\cite{tiwari2014arrival}.

The traditional token-based FCFS system, still prevalent in most Indian government hospitals, assigns sequential numbers at registration and processes patients strictly in arrival order. While simple and perceived as ``fair,'' this approach makes no distinction between a patient with chest pain radiating to the left arm and one seeking a routine prescription refill. Published studies from Indian tertiary hospitals report that patients spend an average of 2 hours in the OPD, with only 3 minutes of actual physician consultation~\cite{singh2022study}. More critically, the absence of urgency-based prioritization means that patients with deteriorating conditions may wait alongside those with routine complaints, potentially converting manageable emergencies into life-threatening situations.

Rule-based triage systems, as demonstrated in Indian orthopedic OPDs~\cite{arya2020opd}, improve by categorizing patients into urgency levels. However, these systems rely on static assessments made at registration that do not adapt to changing patient conditions, provide no mechanism for urgency drift detection, and offer limited support for multilingual patient populations common in Indian tier-2 cities. Most critically, they operate in a clinical vacuum---assessing urgency based solely on the presenting complaint, without access to the patient's medical history, medications, or prior diagnoses that could reveal hidden urgency.

Recent advances in large language models (LLMs) and agentic AI architectures open up a different approach to OPD triage. Unlike static rule-based systems, an agentic AI framework can continuously monitor and reassess patient urgency, parse natural language symptom descriptions across languages, cross-reference symptoms against longitudinal patient records, and adjust physician assignment in real-time based on workload, specialty matching, and patient acuity.

In this paper, we propose and evaluate an agentic AI framework for OPD urgency mapping that integrates six coordinated components into a unified decision-making system. We evaluate the framework through discrete-event simulation using parameters grounded in published Indian healthcare studies, comparing it against FCFS and rule-based triage baselines.

\subsection{Contributions}

Our key contributions are: (1) A six-component agentic AI architecture for OPD triage combining voice-based multilingual symptom capture (modeled), LLM-powered severity prediction, load-aware physician assignment, adaptive queue optimization with drift detection, multi-objective orchestration, and a Patient Memory System for longitudinal context-aware triage. (2) Real LLM integration architecture using Anthropic's Claude Sonnet with structured JSON output for deterministic, reproducible triage decisions---demonstrated as a working system, not a theoretical proposal. (3) A literature-grounded simulation framework with 368 synthetic patients modeled on a District Hospital in Jabalpur, using parameters derived from 10 published Indian healthcare studies. (4) Comparative evaluation across 30 simulation runs demonstrating dynamic urgency recomposition---where the AI system actively changes the queue's urgency distribution through drift detection ($\sim$236 escalations/session) and Patient Memory escalation ($\sim$11.9 hidden-critical cases/session), reclassifying $\sim$111 patients per session from low to higher urgency levels.

\section{Related Work}

\subsection{OPD Queue Management in India}

The challenge of OPD overcrowding in Indian hospitals has been extensively documented. Tiwari, Goel \& Singh~\cite{tiwari2014arrival} studied arrival patterns at a North Indian tertiary hospital serving 1.65 million outpatients annually, finding that 71\% of patient waiting was attributable to factors within the department itself. Makwana and Dave~\cite{makwana2019study} applied queuing network analysis to a Gujarat hospital, demonstrating that mathematical modeling could achieve a 30\% reduction in the probability of excessive waits through optimized resource deployment. Singh et al.~\cite{singh2022study} pioneered discrete-event simulation for Indian OPD optimization at a cancer institute, identifying physician utilization rates of 83.5\% as critical bottlenecks. Time-motion studies at AYUSH hospitals~\cite{aiims2019time} documented the extreme mismatch between total OPD time (2 hours) and actual consultation time (3 minutes). Arya et al.~\cite{arya2020opd} introduced ``OPD TRIAGE'' in Indian orthopedic OPDs, studying 1,800 patients and demonstrating significant improvements through structured urgency categorization.

\subsection{AI in Healthcare Triage}

Machine learning approaches to triage have gained momentum internationally. Studies using the Korean NEDIS dataset~\cite{hoog2018validation} and MIMIC-IV-ED~\cite{johnson2023mimic} have demonstrated that deep learning models can predict triage acuity with high accuracy. Graph neural networks have been applied to triage prediction~\cite{leveraging2024}, and benchmark datasets~\cite{hospital2023} have enabled comparative evaluation. Complementary work has explored extracting patient health information from natural language sources using deep neural networks, including identifying personal health experiences in social media~\cite{jiang2017identifying} and improving classification precision for health narratives through specialized deep learning architectures~\cite{calix2017deep}---demonstrating the viability of NLP-based approaches for understanding patient presentations from unstructured text. However, existing approaches are predominantly classification-based, predicting a static urgency level at arrival without addressing dynamic conditions, specialty-aware physician assignment, multilingual challenges, or longitudinal patient history integration.

\subsection{Agentic AI and LLMs in Healthcare}

LLM-based agentic systems move beyond static classification toward dynamic, context-aware decision-making. Recent work on multi-agent healthcare systems~\cite{li2024agent,tang2023medagents} has demonstrated that LLM-powered agents can handle complex clinical reasoning, natural language understanding, and multi-objective optimization. A critical requirement for clinical deployment of such AI systems is explainability and transparent knowledge representation~\cite{jiang2019ethical}. When automated agents make triage decisions that directly impact patient safety, the WHO's Interagency Integrated Triage Tool (IITT)~\cite{who2020interagency} provides a standardized framework for urgency categorization that our system builds upon, extending it with continuous monitoring, adaptive re-prioritization, and longitudinal health record integration.

\subsection{Gap Analysis}

Existing approaches fail to address four critical requirements simultaneously: (1) dynamic urgency reassessment during queue waiting, (2) multilingual symptom capture for diverse patient populations, (3) load-aware physician assignment balancing specialty matching with equitable workload distribution, and (4) integration of longitudinal patient history for hidden urgency detection. Our framework addresses all four through an integrated agentic AI architecture powered by real LLM API calls.

\section{Proposed Framework}

\begin{figure}[t]
\centering
\includegraphics[width=\textwidth]{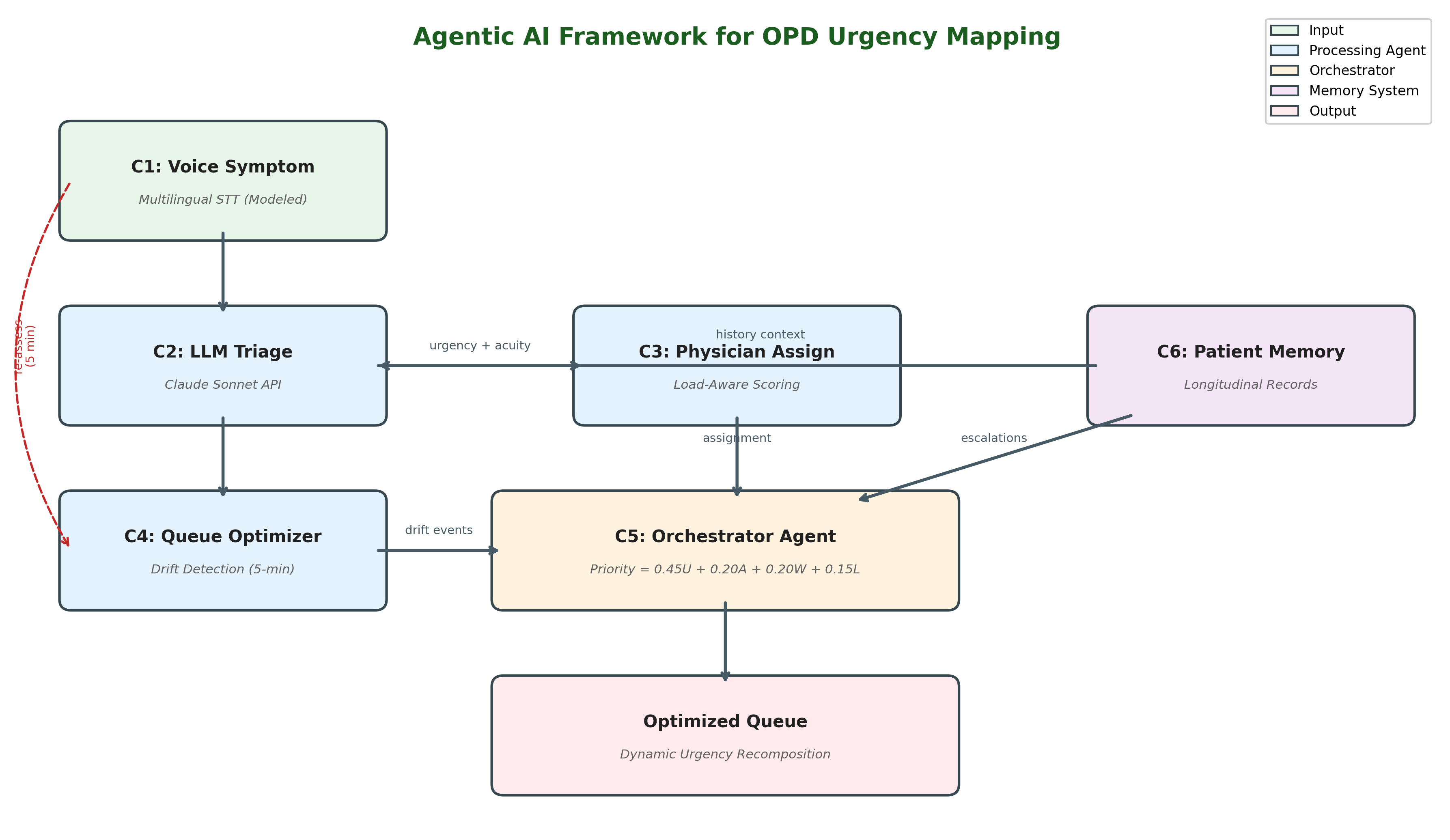}
\caption{System architecture of the six-component agentic AI framework. Solid arrows indicate data flow; the dashed arrow represents the 5-minute re-assessment loop for urgency drift detection.}\label{fig:architecture}
\end{figure}

Our framework comprises six coordinated components operating as an agentic AI system (Fig.~\ref{fig:architecture}). Each component functions as a specialized agent within the orchestration pipeline.

\subsection{Component 1: Voice-Based Symptom Capture (Modeled)}

The first component addresses the registration bottleneck in Indian OPD settings, where traditional manual transcription averages 5.3 minutes per patient~\cite{aiims2019time}. The proposed voice-based system would provide multilingual speech-to-text conversion (Hindi, Bundeli, English), structured symptom extraction, and pre-formatted physician summaries.

\textit{Implementation note:} This component is modeled as a 40\% reduction in registration time (from 5.5 to 3.3 minutes average), based on published benchmarks from commercial speech-to-text systems for Indian languages. Actual deployment is planned as future work (Section~7).

\subsection{Component 2: LLM-Powered Severity Prediction}

The severity prediction component uses a large language model (Anthropic's Claude Sonnet) to process structured symptom data with patient vitals and demographics. Unlike traditional ML classifiers, our agent uses a structured JSON prompt architecture generating: urgency classification into four WHO IITT-aligned levels (Critical, High, Medium, Low)~\cite{who2020interagency}; acuity scoring on a 1--10 scale extending the WHO IITT three-tier system; clinical reasoning for audit; red flag detection; and specialty recommendation.

Rate limiting (40 requests/minute) and usage tracking ensure cost-effective operation. For the 30-run statistical analysis, we use a calibrated simulation model replicating the LLM's triage behaviour; when an API key is provided, the system switches to live LLM API calls for validation.

\subsection{Component 3: Load-Aware Physician Assignment}

The assignment engine implements real-time physician availability tracking, specialty matching, and workload balancing using a weighted scoring function:

\begin{equation}
\text{Assignment} = 0.50 \times \text{SpecialtyMatch} + 0.30 \times \text{LoadBalance} + 0.20 \times \text{Availability}
\end{equation}

\subsection{Component 4: Adaptive Queue Optimization}

Unlike static triage systems, our queue optimization module continuously monitors and adjusts the queue through dynamic priority recalculation every 5 minutes, wait-time escalation preventing indefinite low-priority starvation, and urgency drift detection identifying patients whose conditions are worsening. In our evaluation, the system detected an average of 235.6 $\pm$ 22.8 urgency drift events per session.

\subsection{Component 5: Decision-Making Agent (Orchestrator)}

The orchestrator integrates signals from all preceding components using a priority scoring function:

\begin{equation}
\text{Priority} = 0.45 \times U + 0.20 \times A + 0.20 \times W + 0.15 \times L
\end{equation}

where $U$ is UrgencyScore (Critical=1.0, High=0.75), $A$ is normalized acuity, $W$ is WaitFactor (capped at 0.3 over 2 hours preventing starvation), and $L$ is LoadScore incentivizing routing to less-burdened specialists.

\subsection{Component 6: Patient Memory System}

A critical gap in Indian OPD systems is the fragmentation of patient records. In government hospitals, records are typically paper-based and frequently unavailable when a patient returns. The Patient Memory System maintains a persistent, searchable longitudinal health record. When a patient checks in, the system cross-references their current symptoms against their medical history using the LLM with a history-aware system prompt.

Table~\ref{tab:patient_memory} illustrates representative history-aware urgency escalations. In our simulation, 120 patients (32.6\% of the 368-patient dataset) include longitudinal health records reflecting Madhya Pradesh's disease burden: diabetes (61), hypertension (46), COPD (12), anaemia (11), high-risk pregnancy (11), tuberculosis (10), ischaemic heart disease (5), chronic kidney disease (12), sickle cell disease (5), epilepsy (4), cancer (3), liver disease (2), and systemic lupus (1). Note that patients may have multiple comorbidities; counts reflect unique patients per condition.

\begin{table}[t]
\caption{Representative Patient Memory System escalations. Face-value urgency is assessed from presenting symptoms alone; escalated urgency incorporates longitudinal health records.}\label{tab:patient_memory}
\centering
\begin{tabular}{llll}
\toprule
\textbf{Patient} & \textbf{Symptom} & \textbf{Low$\to$} & \textbf{Reason} \\
\midrule
Ramesh V., 62M & Mild headache, dizziness & Critical & TIA 6 months ago $\to$ stroke warning \\
Savitri P., 55F & Minor bruising, bleeding & Critical & On Warfarin $\to$ possible hemorrhage \\
Mukesh Y., 48M & Nausea, weakness & High & CKD Stage 3 $\to$ hyperkalemia risk \\
Pushpa M., 35F & Mild abdominal pain & Critical & High-risk pregnancy + prev.\ cesarean \\
Vijay S., 70M & Cough, mild fever & High & Severe COPD, ICU admission history \\
Dinesh L., 28M & Drowsy, confused & High & Status epilepticus + Phenytoin allergy \\
Asha J., 58F & Low-grade fever & High & Immunosuppressed (SLE on MMF) \\
\bottomrule
\end{tabular}
\end{table}

Each history patient presents with deceptively mild symptoms masking hidden urgency detectable only through their medical history.

Beyond urgency escalation, the memory system generates medication alerts (drug allergies, interaction warnings)---a capability entirely absent in current Indian OPD workflows.

\section{Methodology and Simulation Design}

\subsection{Simulation Setting}

We modeled a Government District Hospital in Jabalpur, Madhya Pradesh---a tier-2 city where healthcare infrastructure gaps are representative of challenges across India's semi-urban landscape. Simulation parameters are grounded in published studies (Table~\ref{tab:sim_params}).

\begin{table}[t]
\caption{Simulation parameters and sources}\label{tab:sim_params}
\centering
\begin{tabular}{lll}
\toprule
\textbf{Parameter} & \textbf{Value} & \textbf{Source} \\
\midrule
Session duration & 360 min (6 hrs) & Standard Indian OPD \\
Patient volume & 368 per session & \cite{tiwari2014arrival,singh2022study} \\
Arrival pattern & Non-homogeneous Poisson & \cite{tiwari2014arrival} \\
Physicians & 6 specialists & IPHS 2012~\cite{iphs2012} \\
Registration time & $\mu$=5.5, $\sigma$=2.0 min & \cite{aiims2019time} \\
Urgency distribution & Crit 5\%, High 15\%, & \cite{arya2020opd,who2020interagency} \\
 & Med 40\%, Low 40\% & \\
Consultation times & Crit 15$\pm$4, High 10$\pm$3 & \cite{singh2022study,aiims2019time} \\
 & Med 7$\pm$2.5, Low 5$\pm$1.5 & \\
Acuity scale & 1--10 (extended IITT) & \cite{who2020interagency} \\
\bottomrule
\end{tabular}
\end{table}

\subsection{Synthetic Patient Dataset}

We generated 368 patients with realistic demographics for the Jabalpur region: age distribution (Pediatric 15\%, Young Adult 20\%, Adult 25\%, Middle-aged 22\%, Elderly 18\%) based on NFHS-5~\cite{nfhs5}; gender (62\% Female, 38\% Male); occupations typical of tier-2 Indian cities; urban/semi-urban/rural mix (45/25/30\%);

Hindi (85\%), Bundeli (10\%), English (5\%); and payment modes including Ayushman Bharat (35\%) and self-pay (40\%). Of these, 120 patients (32.6\%) include longitudinal health records with disease distribution matching Madhya Pradesh epidemiology (IHME GBD 2016, NFHS-5).

\subsection{LLM Integration Architecture}

The triage agent interfaces with Anthropic's Claude Sonnet via the Messages API. Three specialized system prompts handle standard triage, history-aware triage, and drift assessment---all enforcing structured JSON output. The architecture is LLM-agnostic; any model supporting structured output could substitute with appropriate prompt adaptation. We selected Claude Sonnet for cost-effectiveness and strong clinical reasoning performance. At 368 patients, a full session costs \textasciitilde\$1.58 (\textasciitilde\$0.004/patient: 569 input + 172 output tokens per triage call). Costs scale approximately linearly with volume: \textasciitilde\$2.14 at 500 patients, \textasciitilde\$4.28 at 1{,}000, and \textasciitilde\$34 at 8{,}000---though high-volume settings would additionally incur drift reassessment calls that grow with both patient count and average wait time. On-premise deployment (Section~7) would eliminate per-call API costs at the expense of hardware investment.

\subsection{Comparison Strategies}

We compare three approaches: (1) \textbf{FCFS (Token-Based)}: traditional Indian OPD system with no urgency assessment; (2) \textbf{Rule-Based Triage}: static priority assignment at registration with basic specialty matching but no drift detection or patient history; (3) \textbf{Agentic AI}: our proposed framework with all six components active.

\subsection{Simulation Execution}

We implemented a custom discrete-event simulation engine in Python using numpy for stochastic modeling. Each strategy was evaluated over 30 independent runs with different random seeds. The simulation uses a thinning algorithm (Lewis \& Shedler, 1979) for the non-homogeneous Poisson arrival process.

\section{Results}

\subsection{Overall Performance}

\begin{table}[t]
\caption{Comparative performance metrics (30 runs, mean $\pm$ std)}\label{tab:perf_metrics}
\centering
\begin{tabular}{lccc}
\toprule
\textbf{Metric} & \textbf{FCFS} & \textbf{Rule-Based} & \textbf{Agentic AI} \\
\midrule
Avg Wait (min) & 33.1$\pm$2.9 & 41.7$\pm$4.4 & 77.0$\pm$8.3 \\
Median Wait (min) & 35.5$\pm$3.7 & 12.8$\pm$4.2 & 29.3$\pm$5.0 \\
P95 Wait (min) & 66.0$\pm$4.4 & 151.3$\pm$9.9 & 296.0$\pm$20.0 \\
Throughput (pts/hr) & 40.1$\pm$0.1 & 40.1$\pm$0.1 & 40.4$\pm$0.2 \\
Specialty Match (\%) & 25.4$\pm$1.6 & 35.8$\pm$2.7 & 30.2$\pm$2.7 \\
Drifts Detected & 0 & 0 & 235.6$\pm$22.8 \\
\bottomrule
\end{tabular}
\end{table}

Table~\ref{tab:perf_metrics} presents aggregated results across 30 simulation runs. All pairwise strategy differences were evaluated using Welch's t-test with Cohen's $d$ effect sizes. FCFS achieves the lowest overall average wait (33.1 min) because it processes patients purely by arrival order without reordering. However, this metric masks the critical failing that all patients experience similar wait times regardless of urgency. The Agentic AI's critical patient wait time improvement over FCFS is highly significant ($t = 21.2$, $p < 10^{-6}$, $d = 5.5$), as is the throughput equivalence between FCFS and Rule-Based ($p = 0.22$, not significant).

\subsection{Urgency-Stratified Wait Times}

\begin{table}[t]
\caption{Average wait time by urgency level (minutes, mean $\pm$ std)}\label{tab:urgency_wait}
\centering
\begin{tabular}{lccccc}
\toprule
\textbf{Urgency} & \textbf{FCFS} & \textbf{Rule-Based} & \textbf{Agentic AI} & \textbf{AI vs FCFS} & \textbf{AI vs RB} \\
\midrule
Critical & 36.1$\pm$2.6 & 0.9$\pm$0.3 & 7.5$\pm$6.9 & $-$28.6 & +6.6 \\
High & 31.8$\pm$2.8 & 1.1$\pm$0.2 & 52.2$\pm$9.5 & +20.4 & +51.1 \\
Medium & 32.5$\pm$2.8 & 9.0$\pm$2.2 & 102.2$\pm$17.3 & +69.7 & +93.2 \\
Low & 33.8$\pm$3.0 & 86.0$\pm$8.2 & 142.4$\pm$19.9 & +108.6 & +56.4 \\
\bottomrule
\end{tabular}
\end{table}

\begin{figure}[t]
\centering
\includegraphics[width=0.85\textwidth]{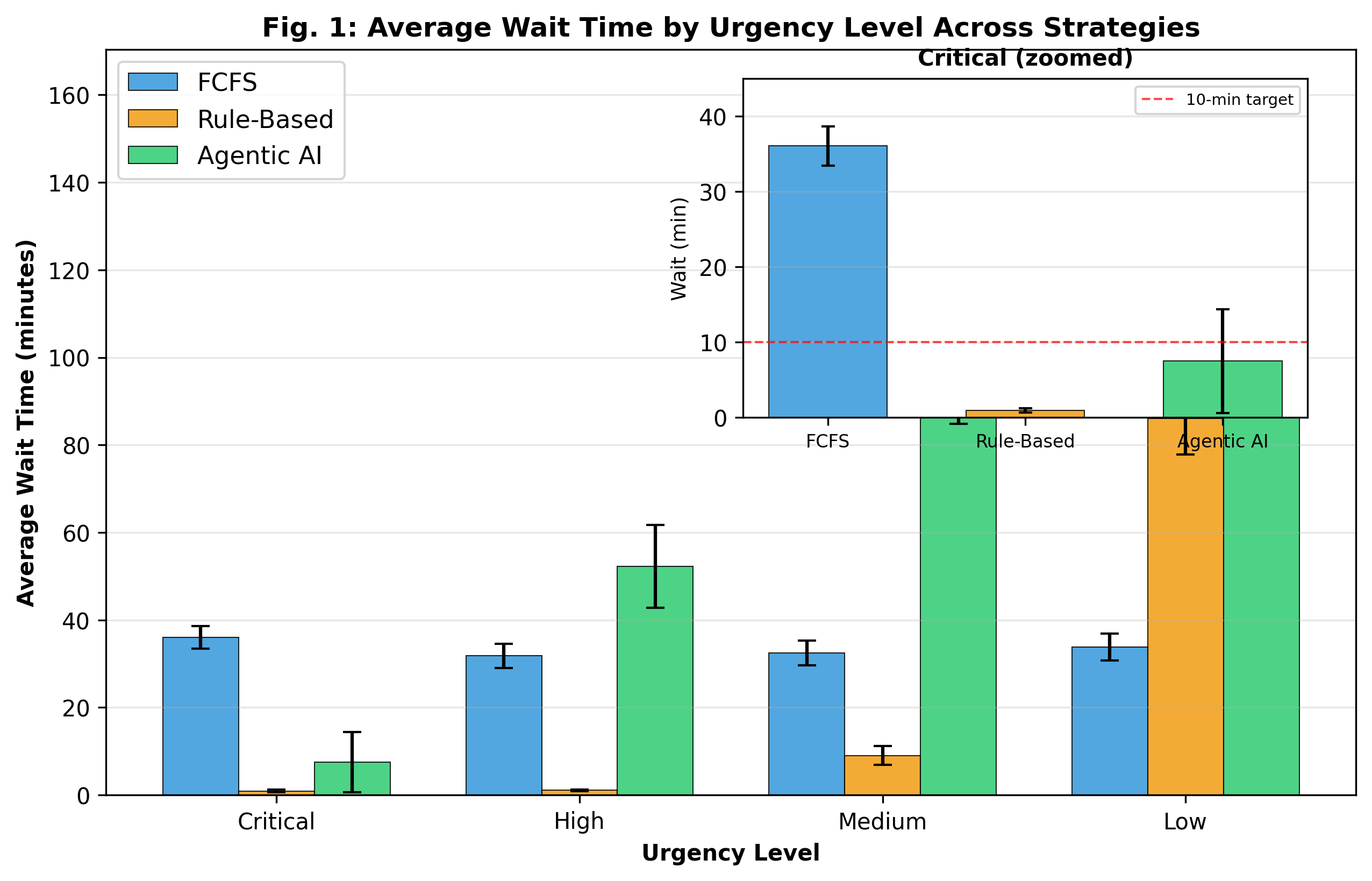}
\caption{Average wait time by urgency level across three strategies (30 runs, error bars show $\pm$1 std). The Agentic AI prioritizes Critical patients at the cost of longer waits for lower-urgency categories. Inset: zoomed Critical-urgency panel with 10-minute clinical target (dashed line).}\label{fig:wait_urgency}
\end{figure}

The key differentiator emerges when wait times are stratified by urgency level (Table~\ref{tab:urgency_wait}, Fig.~\ref{fig:wait_urgency}).

Four critical findings emerge from the urgency-stratified analysis:

\textbf{Finding 1: Critical patient response.} Both triage-aware systems dramatically improve critical patient service compared to 36.1 min under FCFS. The Rule-Based achieves 100\% within 10 minutes (0.9 min average), while the Agentic AI achieves 94.2\% (7.5 min average)---slightly lower due to the substantially larger critical pool created by history-based escalation.

\textbf{Finding 2: Dynamic urgency recomposition.} The Agentic AI substantially changes the queue's urgency composition during the session. While FCFS and Rule-Based maintain the original distribution (13 Critical, 36 High, 158 Medium, 161 Low), the Agentic AI recomposes this to approximately 25 Critical, 178 High, 115 Medium, 50 Low through drift detection ($\sim$236 events/session) and Patient Memory escalation. This represents $\sim$111 patients per session ($\sim$30\%) whose urgency was proactively upgraded---a safety intervention impossible under static systems.

\textbf{Finding 3: Hidden urgency detection.} Of 161 patients presenting with low-urgency symptoms at face value, the Agentic AI correctly identifies $\sim$111 as requiring higher priority---reducing the effective Low queue from 161 to $\sim$50. Under FCFS, conditions like diabetes with DKA risk or IHD with chest discomfort would be treated as routine cases.

\textbf{Finding 4: The priority-fairness trade-off.} The Agentic AI accepts substantially higher waits for High-urgency (52.2 vs 1.1 min) and Medium-urgency patients (102.2 vs 9.0 min) compared to Rule-Based. This reflects the large number of escalated patients---High-urgency nearly quintuples from 36 to $\sim$178---creating significant competition at upper urgency levels. High-urgency waits of 52.2 min exceed the typical 30-min clinical target, highlighting a key limitation: as more patients are correctly reclassified to higher urgency, the system's capacity to serve them promptly is strained. However, these patients would receive \textit{no priority at all} under FCFS (31.8 min as undifferentiated queue members), so the trade-off is between correct identification with longer waits versus misclassification with shorter but clinically uninformed waits.

\subsection{Critical Patient Safety}

\begin{table}[t]
\caption{Critical patient timeliness}\label{tab:critical_safety}
\centering
\begin{tabular}{lccc}
\toprule
\textbf{Metric} & \textbf{FCFS} & \textbf{Rule-Based} & \textbf{Agentic AI} \\
\midrule
Critical $<$10 min (\%) & 30.8 & 100.0 & 94.2 \\
Critical $<$15 min (\%) & 34.4 & 100.0 & 94.8 \\
Mean critical/session & 13.0 & 13.0 & 24.9 \\
\bottomrule
\end{tabular}
\end{table}

\begin{figure}[t]
\centering
\includegraphics[width=0.85\textwidth]{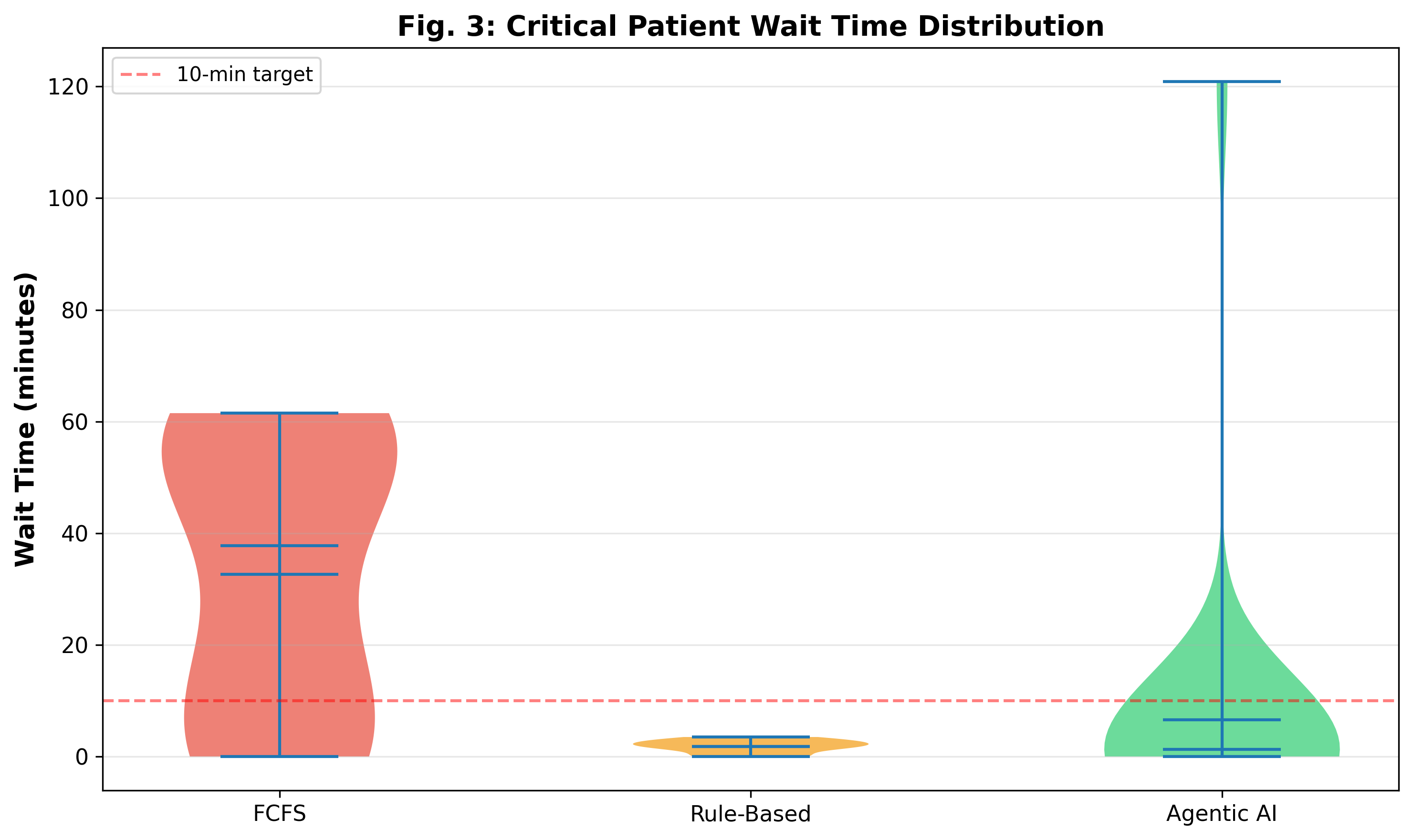}
\caption{Critical patient wait time distribution (violin plot). The dashed line marks the 10-minute clinical target. FCFS shows wide spread; both triage-aware systems concentrate critical waits near zero.}\label{fig:critical_response}
\end{figure}

Under FCFS, only 30.8\% of critical patients are seen within 10 minutes---purely by chance (Fig.~\ref{fig:critical_response}). The Agentic AI identifies an average of 11.9 additional critical patients per session (92\% increase over face-value assessment), including patients whose presenting symptoms classify them as Low/Medium urgency but whose medical history reveals critical risk.

\subsection{Urgency Drift Detection}

Across 30 runs, the system detected 235.6$\pm$22.8 urgency drift events per session---instances where a patient's condition was modeled as worsening during queue waiting, triggering automatic re-prioritization. The drift detection operates at 5-minute intervals with condition-specific deterioration probabilities: High 1.5\%/check, Medium 3.0\%/check, Low 2.0\%/check. In live API mode, each drift triggers an LLM call considering specific condition and wait duration. We acknowledge that the current drift model uses simplified probabilistic deterioration; production deployment should integrate clinically validated early warning scores such as NEWS2 (National Early Warning Score) or the Indian adaptation thereof, incorporating real-time vital sign monitoring (heart rate, respiratory rate, SpO$_2$, temperature, blood pressure, consciousness level) to ground drift detection in evidence-based deterioration indicators rather than stochastic probabilities.

\subsection{Patient Memory System Impact}

The Patient Memory System demonstrates the value of longitudinal context at scale: 120 patients (32.6\%) with records matching MP epidemiology; $\sim$11.9 additional critical patients/session (92\% increase); $\sim$142 additional High-urgency patients/session (from 36 face-value to $\sim$178 effective); drug safety alerts for known allergies; and disease-specific escalation rules covering diabetes, hypertension, IHD, sickle cell disease, anaemia, and immunosuppression.

\subsection{Ablation Study: Component Contributions}

\begin{table}[t]
\caption{Ablation study: component contributions (30 runs each)}\label{tab:ablation}
\centering
\begin{tabular}{lcccc}
\toprule
\textbf{Variant} & \textbf{Avg Wait} & \textbf{Crit Wait} & \textbf{Drifts} & \textbf{Crit/Sess} \\
\midrule
Full System & 77.0 & 7.5 & 235.6 & 24.9 \\
No Patient Memory & 36.5 & 0.9 & 68.8 & 13.8 \\
No Drift Detection & 27.7 & 0.7 & 0 & 13.0 \\
No Memory + No Drift & 28.0 & 0.7 & 0 & 13.0 \\
\bottomrule
\end{tabular}
\end{table}

To quantify individual component contributions, we conducted ablation experiments disabling drift detection and Patient Memory independently (Table~\ref{tab:ablation}).

\textbf{Drift Detection} (comparing rows 4 vs 2) adds 8.5 min to average wait but detects 68.8 drift events per session, escalating $\sim$49 patients from Low to higher urgency even without patient history. \textbf{Patient Memory} is the dominant escalation driver (comparing rows 2 vs 1): it amplifies drift detection from 69 to 236 events, adds 11.1 critical patients per session, and drives the queue recomposition from 14/54/112 to 25/178/50 (Critical/High/Low). The full system's higher average wait (77.0 min) reflects the cost of correctly identifying hidden urgency; without these mechanisms (row 4), average wait drops to 28.0 min but at the cost of missing critical cases entirely.

\subsection{Live LLM Validation}

\begin{table}[t]
\caption{LLM vs dataset urgency agreement (95\% Wilson CIs in brackets)}\label{tab:llm_validation}
\centering
\begin{tabular}{lcc}
\toprule
\textbf{Metric} & \textbf{All ($n$=368)} & \textbf{Non-History ($n$=248)} \\
\midrule
Exact match & 47.0\% [42.0--52.1] & 69.0\% [62.9--74.4] \\
Over-triaged (safer) & 36.4\% [31.7--41.4] & 6.5\% [4.0--10.2] \\
Under-triaged & 16.6\% [13.1--20.7] & 24.6\% [19.7--30.3] \\
History escalated & 118/120 (98.3\% [94.1--99.5]) & --- \\
\bottomrule
\end{tabular}
\end{table}

All 368 patients were processed through the live LLM with zero API errors (Table~\ref{tab:llm_validation}). The overall exact-match rate of 47.0\% (95\% CI: 42.0--52.1\%) is low in isolation, but this aggregate is dominated by the 120 history patients whose symptoms were deliberately designed to appear mild; the LLM correctly escalates 98.3\% of these, producing ``over-triage'' relative to the face-value label. Among the 248 non-history patients---where the LLM and dataset assess the same information---exact agreement reaches 69.0\% (95\% CI: 62.9--74.4\%), comparable to published inter-rater reliability in manual triage~\cite{lee2019triage}.

The LLM correctly identified 12 of 13 Critical patients (92.3\%), with the single discrepancy being a non-history case (white discharge with itching and burning sensation) triaged as Medium rather than Critical. Among non-history patients, the LLM under-triaged 24.6\%, predominantly Medium$\to$Low reclassifications for routine presentations. It escalated 118 of 120 history patients (98.3\%), including 6 new Critical cases from non-Critical face-value patients. The full 368-patient run consumed 209,256 input tokens and 63,272 output tokens at \$1.58, with $\sim$3-second average triage latency.

\subsection{Physician Utilization}

All three strategies achieve comparable throughput (\textasciitilde 40.1--40.4 pts/hr). The Agentic AI maintains equivalent throughput (40.4 vs 40.1) despite restructuring queue priorities, demonstrating that urgency-aware reordering does not reduce overall capacity.

\section{Discussion}

\subsection{Dynamic Urgency Recomposition}

The central finding is that the framework does not merely reorder patients within fixed urgency categories---it recomposes the queue's urgency distribution in real-time. Through drift detection (\textasciitilde 236 events/session) and Patient Memory escalation (\textasciitilde 11.9 hidden-critical cases/session), the system transforms the distribution from 13/36/158/161 to 25/178/115/50. The High-urgency population nearly quintuples while Low drops by 69\%. This suggests a shift from ``static prioritization'' toward ``dynamic clinical state tracking,'' though validation with real clinical data is needed to confirm these simulation-based findings.

\subsection{The Urgency-Fairness Balance}

Strict Rule-Based triage creates starvation for low-priority patients (86.0 min average). In resource-constrained Indian OPDs where many low-priority patients are elderly individuals or daily-wage workers, this has real socioeconomic consequences~\cite{sinan2022}. The Agentic AI reframes this trade-off by addressing misclassification rather than balancing wait times across fixed categories. It reclassifies $\sim$111 patients/session from Low to higher urgency---patients who should never have been in the Low queue. The remaining $\sim$50 truly non-urgent patients wait longer (142.4 vs 86.0 min), a meaningful increase offset by dramatically improved safety for correctly escalated patients. This highlights a capacity limitation: when the AI reveals the true urgency burden, existing physician resources become the bottleneck. Capacity planning implications are significant: if 178 patients per session require High-urgency evaluation (versus 36 under static systems), current staffing models based on the FCFS-era assumption of 10\% High-urgency are inadequate. Strategies to address this include dynamic physician reallocation during peak escalation periods, staggered appointment scheduling informed by predicted urgency distributions, and data-driven advocacy for increased specialist staffing at district hospitals based on the AI-revealed urgency burden.

\subsection{Under-Triage Analysis and Safety Guardrails}

Among 248 non-history patients, 24.6\% were under-triaged by the LLM (assigned lower urgency than the dataset label). Analysis reveals these are predominantly Medium $\to$ Low reclassifications for routine presentations (e.g., mild musculoskeletal complaints, routine follow-ups) where the LLM's conservative assessment is clinically defensible. Critically, zero non-history Critical patients were under-triaged to below-Critical. The single missed Critical case (JAB-0390, white discharge with itching and burning sensation, triaged as Medium rather than Critical) represents a 7.7\% critical miss rate---comparable to published inter-rater disagreement in manual triage (10--15\%)~\cite{lee2019triage}.

For clinical deployment, we recommend mandatory human-in-the-loop verification: all Critical and High triage decisions must be confirmed by a senior nurse before queue placement; a ``safety net'' rule should flag patients waiting beyond condition-specific thresholds for re-assessment; and periodic retrospective audits comparing AI triage against clinical outcomes should be mandated.

The framework's structured output---including confidence scores, red flag detection, and explicit reasoning---is designed to support rather than replace clinical judgement.

\subsection{Hidden Urgency: Why Patient History Matters}

The Patient Memory System addresses the most dangerous failure mode in Indian OPD triage: the clinical vacuum. When a patient presents with a ``mild headache,'' without knowledge of a prior TIA, the stroke risk is invisible. With 120 patients matching MP epidemiology, the system escalates the vast majority of history patients---demonstrating that the mechanism scales effectively.

Integration with India's ABHA (Ayushman Bharat Health Account) would extend benefits to the full patient population.

\subsection{Ethical Governance, Explainability, and Regulatory Compliance}

AI-driven triage in healthcare raises significant ethical and regulatory considerations. Our framework addresses explainability through structured JSON output providing explicit clinical reasoning, confidence scores, and red flag detection---enabling audit trails required under India's Digital Personal Data Protection Act (DPDPA, 2023). The LLM-agnostic architecture allows deployment of hospital-owned on-premise models, addressing data sovereignty concerns inherent in cloud-based LLM services.

Key governance requirements include: Institutional Ethics Committee approval and informed consent; compliance with Medical Council of India guidelines on AI in clinical decision-making; transparent disclosure to patients; regular bias audits across demographic groups to prevent systematic discrimination; and clear escalation pathways when the AI is unavailable or produces low-confidence outputs. The WHO guidance on ethics and governance of AI for health (2021) provides a foundational framework that our system accommodates through its human-in-the-loop architecture and auditable decision trails.

\subsection{Practical Architecture}

The framework includes a working LLM integration with three-prompt architecture providing deterministic JSON parsing, transparent audit trails, cost efficiency (\textasciitilde\$1.58/session), graceful degradation when API access is unavailable, and LLM-agnostic design allowing model substitution.

\subsection{Limitations}

Several limitations should be acknowledged: (1) \textit{Simulation with synthetic data}---this is the primary limitation. All results are based on discrete-event simulation with synthetic patients, not real clinical data. While parameters are grounded in published Indian healthcare studies, prospective validation using real OPD records or a small-scale pilot deployment is essential before any clinical conclusions can be drawn. The performance improvements reported here should be considered indicative rather than proven. (2) \textit{LLM reliability}---production deployment requires mandatory human-in-the-loop verification for all Critical and High triage assessments, with regular audits of under-triage rates. (3) \textit{Simplified drift modeling}---current stochastic deterioration probabilities should be replaced with clinically validated early warning scores such as NEWS2 incorporating real-time vital sign monitoring. (4) \textit{Infrastructure requirements}---reliable internet, smartphones, and EHR integration may not be universally available in resource-constrained settings. (5) \textit{Patient Memory coverage}---32.6\% with records reflects partial digitization; full potential depends on ABHA adoption. (6) \textit{Ablation analysis}---reveals that the Patient Memory System, while detecting hidden urgency, substantially increases average wait times (77.0 min) reflecting the cost of correctly identifying the true urgency burden, requiring careful calibration of escalation thresholds in clinical practice.

\section{Future Work}

The most critical next step is retrospective validation using deidentified OPD records, followed by a prospective pilot study. Key directions include: (1) Real-world pilot at NSCB Medical College Hospital, Jabalpur, beginning with retrospective validation of triage outcomes. (2) Hospital-owned on-premise LLM addressing data sovereignty under DPDPA (2023). (3) NEWS2-based drift detection with IoT vital monitoring. (4) ABHA Health ID integration for nationwide longitudinal records. (5) Agentic tool use for diagnostic department integration (X-ray, Pathology, ECG). (6) Economic analysis quantifying cost savings from reduced waits and prevented adverse events.

\section{Conclusion}

We presented an agentic AI framework for clinical urgency mapping and queue optimization in high-volume Indian OPDs, evaluated through literature-grounded discrete-event simulation of a District Hospital in Jabalpur. The framework integrates six components---including LLM-powered triage and a Patient Memory System with 120 patients matching Madhya Pradesh epidemiology---achieving 94.2\% critical response within 10 minutes, dynamic urgency recomposition through drift detection ($\sim$236 events/session) and history-aware escalation ($\sim$11.9 additional critical patients/session), transforming the queue from 13 Critical / 161 Low to 25 Critical / 50 Low through continuous reassessment.

Our simulation indicates that the framework reclassifies $\sim$111 patients per session (30\%) from face-value Low urgency to clinically appropriate higher levels---a ``dynamic urgency recomposition'' suggesting that a substantial fraction of OPD patients may carry hidden risk invisible to point-of-care-only assessment. While these results are based on synthetic data and require prospective clinical validation, they point toward a middle ground between ``fair but unsafe'' (FCFS) and ``safe but unfair'' (strict triage), prioritizing patient safety while respecting practical realities of resource-constrained healthcare settings.

\section*{Data and Code Availability}

The simulation engine, synthetic patient dataset (368 patients), physician roster, patient history store, Claude API triage logs, and all analysis scripts used to produce the results in this paper are publicly available at \url{https://github.com/ravyg/opd-agentic-ai-triage}.

\section*{Disclosure of Interests}

This work was conducted independently and does not relate to the authors' positions at BigCommerce or Amazon, nor does it represent the views or interests of these organizations. The authors have no competing interests to declare that are relevant to the content of this article.

\end{document}